\documentstyle[epsfig]{aipproc}

\begin{document}
\title{Black Holes and Galaxy Metamorphosis.}

\author{J. Kelly Holley-Bockelmann}
\address{Case Western Reserve University\\
Cleveland, Ohio 44106\\}

\maketitle

\begin{abstract}
Supermassive black holes
can be seen as an agent of galaxy transformation.
In particular, a supermassive black hole can cause a triaxial galaxy to evolve 
toward axisymmetry by inducing chaos in centrophilic orbit
families. This is one way in which a single supermassive 
black hole can induce large-scale changes in the structure
of its host galaxy -- changes on scales far larger than the
Schwarzschild radius ($O(10^{-5}) \rm{ pc}$) and the  
radius of influence of the black hole ($O(1)-O(100) \rm{ pc}$).  
\end{abstract}

\section*{Introduction}
Observations are beginning to conclude that massive central black holes 
are a natural part of elliptical galaxy centers \cite{kormendy00,richstone98}.
In fact, best-fit models of 
black hole demography indicate that approximately $97\%$ of 
ellipticals harbor a massive black hole.
Black hole mass seems to be correlated with the
host bulge potential; current dynamical estimates of the best 
galactic black hole candidates 
have yielded masses on the order of $0.005 M_{\rm bulge}$\cite{kormendyrichstone95,magorrian98,vandermarel99}.
There also seems to be
a strong correlation between black hole mass and global velocity
dispersion, implying that the Fundamental Plane exists even in
the four-dimensional space described by ($\log M_{\rm BH}, \log L, 
\log \sigma_e, \log R_e$) \cite{Gebhardt00,merritt00}. It appears, then, 
that black hole formation and the evolution of the host galaxy 
may be deeply connected. This proceeding explores one way in which 
black holes may drastically change the structure of an elliptical 
galaxy over large scales: by inducing axisymmetry in a triaxial galaxy.
It appears, then, that a supermassive black hole's impact on the structure and 
subsequent evolution of its host galaxy is both dramatic and far-reaching.

\section*{Black Holes and Triaxial Galaxies}

\begin{figure}[b!]
\centerline{\epsfig{file=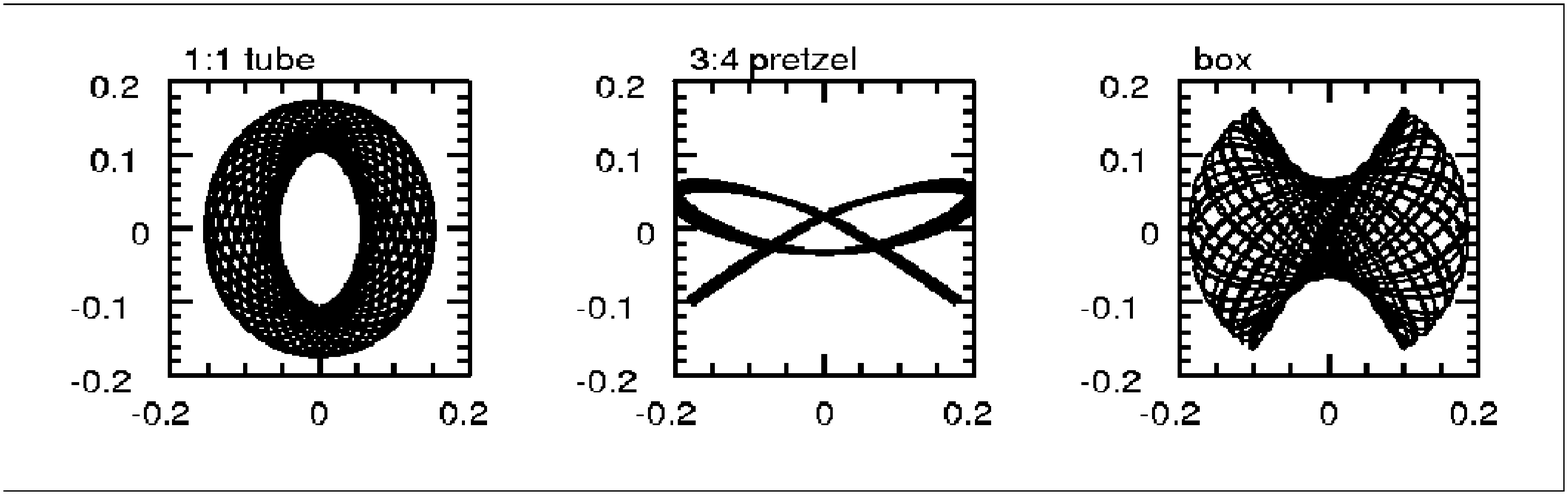,height=2.in,width=7in}}
\vspace{10pt}
\caption{Planar Orbits in Triaxial Potentials.}
\label{fig1}
\end{figure}

There is evidence, both observationally and theoretically, that elliptical
galaxies are at least mildly triaxial in shape \cite{bak00,dubinski91}. Even a mildly triaxial
galaxy will generate entirely different orbit families than are present in
a spheroid. In particular, there are a rich variety 
of regular box and boxlet orbits that are centrophilic and 
comprise the backbone of the galaxy (Figure \ref{fig1}). These centrophilic orbits
can be driven chaotic with the introduction of a supermassive black hole \cite{norman85,gerhard85,merritt97,quinlan98,vallurimerritt98,holley01}. 
And, since chaotic orbits will eventually fill all available phase space, 
the time-averaged shape of a chaotic orbit is spherical. Hence, 
the destruction of these centrophilic orbits breaks the backbone of the
triaxial model and the system evolves toward axisymmetry. This effect has been shown in numerous computational and analytic studies\cite{norman85,gerhard85,merritt97,quinlan98,vallurimerritt98,holley01}. In fact, the main
controversies are whether the galaxy becomes axisymmetric 
locally or globally, and whether the transformation occurs in a
few crossing times or over many Hubble times.

If the transformation is rapid and global, there are important implications
for elliptical galaxy evolution. For example, one possible difference between
an intrinsically bright elliptical (which is thought to be more triaxial) 
and a faint elliptical is that the faint elliptical, with its shorter
crossing time, has had more interactions with the black hole and is thus
more dynamically evolved \cite{vallurimerritt98}. Secondly, the 
black hole/bulge mass relation can be explained in terms of 
galaxy evolution \cite{vallurimerritt98}. In this scenario, 
spiral galaxies begin as gas-rich disks with
a small triaxial bulge. Since triaxiality supports box orbits, 
gas can flow radially inward along these orbits,  
which can rapidly funnel matter into a black hole. The black hole grows
until a critical black hole mass of $M_\bullet = 0.02M_{\rm gal}$, 
which breaks triaxiality and strongly curtails the gas inflow. 
Subsequent disk-disk merging can create a elliptical galaxy, and black hole 
feeding ensues in this larger triaxial bulge until the critical black
hole mass is achieved. In both types of galaxies, the process is the
same: once the black hole mass fraction is large enough to disrupt
box orbits, gas inflow is sharply diminished.

However, most self-consistent N-body simulations which have
studied this effect
have employed astrophysically unrealistic galaxy models \cite{norman85,quinlan98}. For example, the models
have been highly flattened, maximally triaxial models, while observations
indicate that ellipticals are most likely mildly triaxial and not 
very flattened. Furthermore, the black hole was often grown within a
flat inner density profile (a 'core') which, while it maximized the change 
in the inner potential, does not reflect the observations of cuspy
inner density profiles in ellipticals.

My collaborators and I \cite{holley01}
have explored the issue of black hole-embedded triaxial galaxies
with more realistic galaxy models. Our models exhibited a range of
moderate triaxialities ($0.25 < T < 0.75$) with mild flattening and cuspy
inner density profiles. They were generated by 'adiabatically squeezing'
a cuspy Hernquist sphere into a stable triaxial figure, and then 
adiabatically growing a central black hole in this model\cite{holley01}.
The models were populated with $N=512,000$ multimass particles to 
accurately reproduce the central density cusp, and the particles
were advanced with a multiple timestep, high-order Hermite integrator
in the SCFcode.

\begin{figure}[b!]
\centerline{\epsfig{file=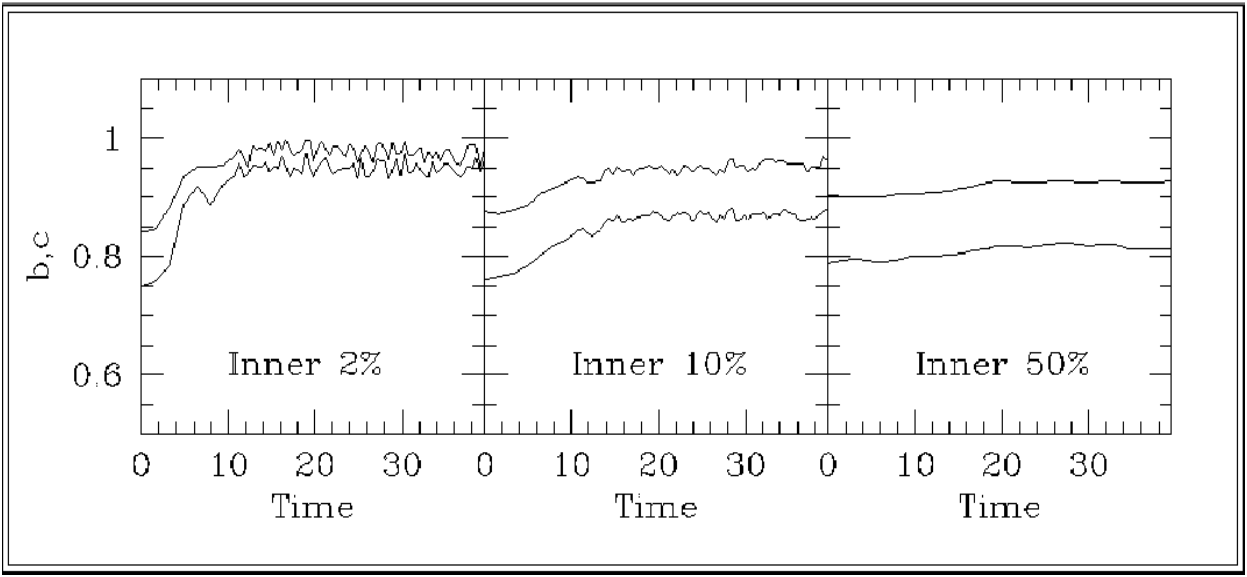,height=3.in,width=6.5in}}
\vspace{10pt}
\caption{The intermediate and minor axes lengths as a function of time 
for particle sets binned by mass in the model. The axes lengths 
are iteratively calculated from the ellipsoidal
density distribution using the moment of inertia tensor. The lack of 
evolution in the axes lengths in the last panel argues for a stable
shape at the half mass radius, in spite of the black hole.}
\label{fig2}
\end{figure}

Figure \ref{fig2} shows the change in the physical structure of a
triaxial model as the black hole is grown. Initally, the model
had axis lengths $a:b:c=1:0.85:0.7$, and a central density 
cusp $\gamma=1$. If this galaxy were to lie on the core and global
Fundamental Planes, the density cusp dictates the absolute magnitude 
($M_v\approx-21.6$), mass ($M_{\rm gal} \approx 2{\rm x}10^{12} M_\odot$), core radius ($r_{\rm core}\approx 150 \rm{ pc}$) and effective radius ($r_{\rm eff} 
\approx 4 \rm{ kpc}$) of the corresponding galaxy.  

\begin{figure}[b!]
\centerline{\epsfig{file=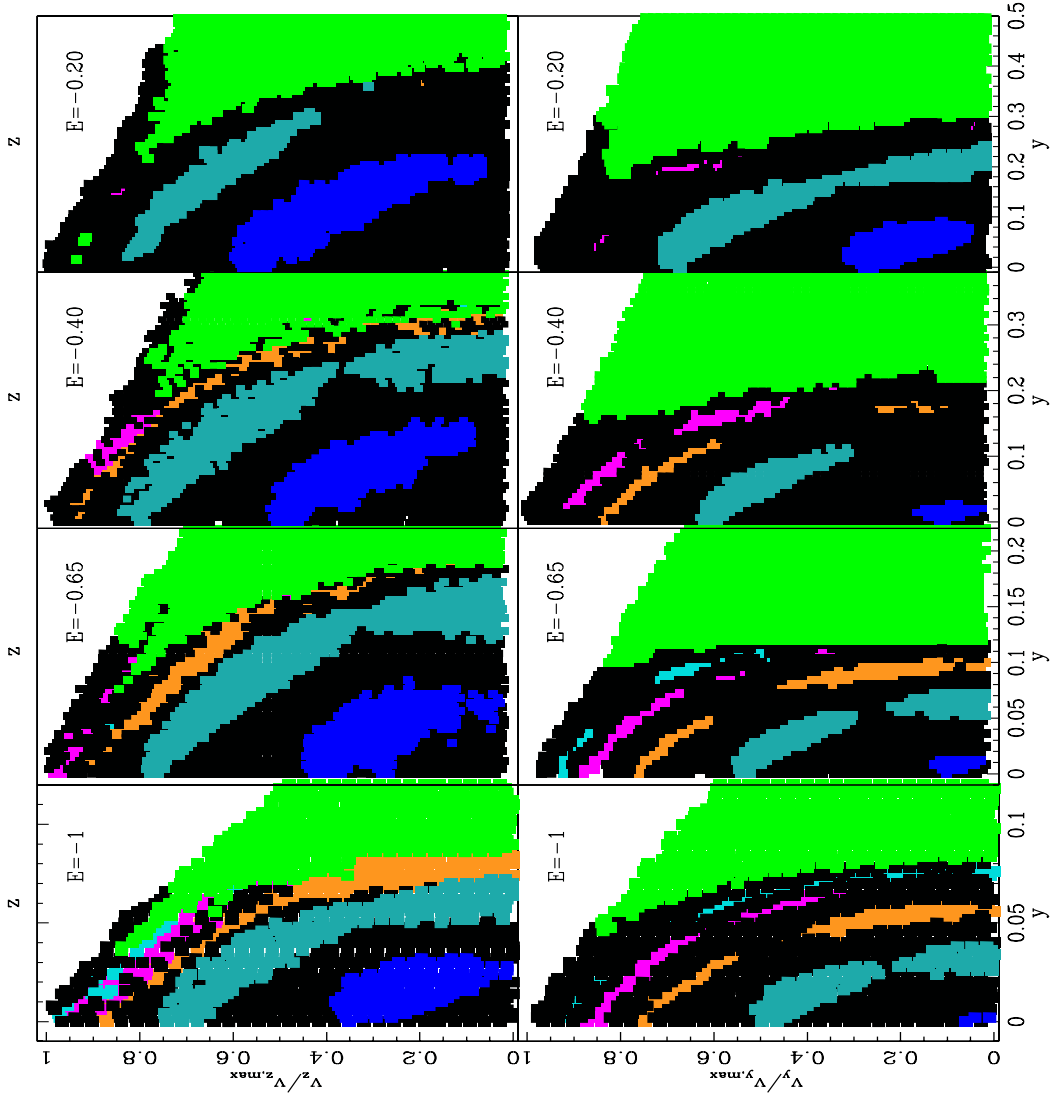, height=6.4in,width=6in}}
\vspace{-25pt}
\caption{Surfaces of section for the initial triaxial model without a central black hole,
plotted for orbital populations of differing binding energies. Top:
Surfaces of section for orbits in the $xz$ plane. Bottom: Surfaces of
section for orbits in the $xy$ plane. Orbits are coded by color --
strongly chaotic:red, loops:green, boxes:black, bananas:yellow, fish:blue, pretzels:aqua, 5:4 resonance:orange,
6:5 resonance:magenta, 7:6 resonance:cyan. We zoom in on the plot's x-axis to show as
many box and boxlets as possible; the orbits outside the limits of the plot 
are tubes. If plotted to the full extent of the x axis, the
boxlet region would comprise $\sim 50\%$ of the most bound panels, and only $\sim 10\%$
of the least bound panels. Notice that there are no strongly chaotic 
regions in this stable triaxial figure.}
\label{fig3}
\end{figure}

Although the final state of this black hole-embedded model is decidedly
not axisymmetric on global scales, it is certainly more round near the
center. We developed an automated orbit analysis routine to determine
if this central roundening was caused by black hole-induced centrophilic 
orbit destabilization.
Figure \ref{fig3} presents xy,xz surfaces of section for the initial state 
of the triaxial potential (ie no black hole). Despite the rich variety of resonant boxlets,
boxes, and tubes, less than $0.2\%$ of the orbits were strongly chaotic
(see \cite{holley01} for details).

After the black hole has grown, the orbital content is quite different. 
In the inner regions, nearly all centrophilic orbits have become 
stochastic, including the population of eccentric loops. However,
for less bound orbits, there are only a scattered few strongly chaotic orbits.
A common interpretation for the lack of chaos in these less bound orbits
is that they have been integrated for far fewer dynamical times than the
highly bound orbits. Thus these outermost orbits have not been exposed to the perturbative
black hole potential enough for substantial chaos to set in.
To test this explanation, we integrated a
subset of the lesser bound box and boxlet orbits ($E=-0.40$) in 
the xy plane for
$\approx 200$ orbital times. Although the percentage of chaotic orbits
in this subset increased from $4\%$ to $71\%$ over the experiment, many stable 
centrophilic orbits existed after
$\approx$ 2 Hubble times, including a large fraction of non-resonant boxes.
A possible explanation is that these weakly bound orbits spend 
very little of the orbit in the inflection region between spherical
potential and the triaxial potential, and are thus not driven 
stochastic at all. This explanation is far different than scenarios
which involve scattering by the black hole, but black hole scattering 
is difficult to envision as the cause of the chaos observed in our models,
since there were many outer orbits that passed quite close to the black
hole, yet remained stable after hundreds of dynamical times.

\section*{Conclusion}

The black hole in our model induced axisymmetry out to nearly 100 parsecs, 
and resulted in a clearly observable change in the shape and structure
of the galaxy. Since the transformation did not take place globally,
it is tempting to say that the black hole mass/bulge mass relation
observed in the current galaxy population is {it not} simply an artifact 
of gas inflow in a more triaxial-shaped progenitor population.
However, it is not immediately clear how the more localized 
axisymmetry we observed would
effect gas inflow and subsequent black hole feeding. While it it true in our
globally triaxial model that gas inflow from
outside the half mass radius would never be entirely cut off, the
behavior of the gas once it 
hits the axisymmetric region requires detailed gas dynamical simulations.
Nonetheless, it is clear that a central supermassive black hole 
causes dramatic and long-lasting
changes in the host galaxy over scales well outside the region in which
it dominates the potential.  

\begin{figure}[b!]
\centerline{\epsfig{file=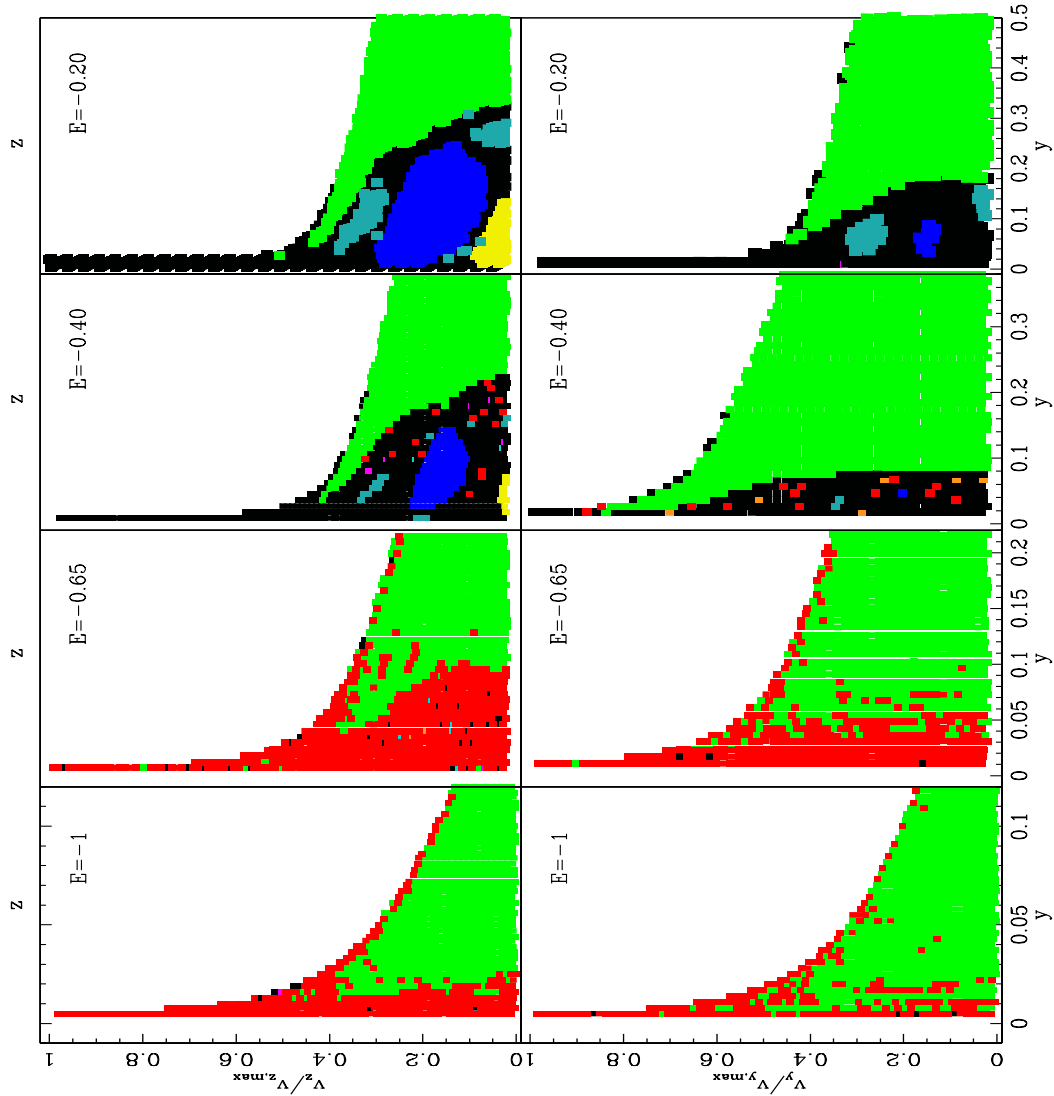, height=6.4in,width=6in}}
\vspace{-25pt}
\caption{Surfaces of section for the final state of the triaxial model with a
central black hole of mass $M_{\bullet}=0.01 M_{\rm gal}$. See Figure \ref{fig3} for explanation of symbols. Notice that the box and boxlet space in the 
inner regions of this model is almost entirely taken over by chaotic orbits,
while the outer region is nearly devoid of strongly chaotic orbits.}
\label{fig4}
\end{figure}

\end{document}